\documentclass[11pt]{article}
\usepackage{hyperref}
\hypersetup{pdfborder = {0 0 0},colorlinks=true,linkcolor=blue,urlcolor=blue,citecolor=blue}
\newcommand{\proglang}{\texttt}
\newcommand{\pkg}{\texttt}
\newcommand{\code}{\texttt}

\usepackage{verbatim}

\newenvironment{CodeInput}%
{\endgraf\verbatim}%
{\endverbatim}
\newenvironment{CodeOutput}%
{\endgraf\verbatim}%
{\endverbatim}
\usepackage[top=1in,bottom=1in,left=1in,right=1in]{geometry}

\usepackage[utf8]{inputenc}
\usepackage[T1]{fontenc}
\usepackage{amsmath,amssymb,array}
\usepackage[round]{natbib}
\usepackage{graphicx}
\usepackage{booktabs}
\usepackage{xcolor}
\usepackage{float}

\usepackage{enumitem}

\usepackage{amsfonts, amssymb, amsmath}
\usepackage{multirow, setspace}
\usepackage[justification=centering]{caption}
\usepackage{subcaption}
\usepackage{dsfont}

\newcommand{\Prob}{\mathbb{P}}
\newcommand{\Indicator}{\mathds{1}}

\newcommand{\Trans}{\top}

\renewcommand{\epsilon}{\varepsilon}

\newcommand{\pOrder}{\mathfrak{p}}
\newcommand{\qOrder}{\mathfrak{q}}
\newcommand{\Ysupp}{\mathcal{Y}}

\newcommand{\bQ}{\mathbf{Q}}
\newcommand{\bR}{\mathbf{R}}
\newcommand{\bS}{\mathbf{S}}

\newcommand{\bX}{\mathbf{X}}

\newcommand{\bb}{\mathbf{b}}
\newcommand{\bc}{\mathbf{c}}

\newcommand{\be}{\mathbf{e}}

\newcommand{\bm}{\mathbf{m}}

\newcommand{\bp}{\mathbf{p}}
\newcommand{\bq}{\mathbf{q}}

\newcommand{\bu}{\mathbf{u}}

\newcommand{\bx}{\mathbf{x}}

\newcommand{\bbeta}{\boldsymbol{\beta}}

\newcommand{\bnu}{\boldsymbol{\nu}}
\newcommand{\bgamma}{\boldsymbol{\gamma}}

\newcommand{\Y}{\mathcal{Y}}
\newcommand{\X}{\mathcal{X}}
\newcommand{\Z}{\mathcal{Z}}
\newcommand{\A}{\mathcal{A}}
\newcommand{\B}{\mathcal{B}}

\newcommand{\M}{\mathcal{M}}

\newcommand{\inv}{{-1}}
\newcommand{\eval}{y, \bx}

\DeclareMathOperator*{\argmin}{argmin}

\newcommand{\sumIN}{\sum_{i=1}^n}

\title{\texttt{lpcde}: Estimation and Inference for Local Polynomial Conditional Density Estimators}
\author{Matias D. Cattaneo\thanks{Department of Operations Research and Financial Engineering, Princeton University.} \and
  Rajita Chandak\thanks{Institute of Mathematics, \'{E}cole Polytechnique F\'{e}d\'{e}rale de Lausanne} \and
  Michael Jansson\thanks{Department of Economics, UC Berkeley.} \and
  Xinwei Ma\thanks{Department of Economics, UC San Diego.}}

\begin{document}

\maketitle

\begin{abstract}
    This paper discusses the \proglang{R} package \pkg{lpcde},
    which stands for local polynomial conditional density estimation.
    It implements the kernel-based local polynomial smoothing methods introduced in
    \cite{Cattaneo-Chandak-Jansson-Ma_2024_Bernoulli}
    for statistical estimation and inference of conditional distributions,
    densities, and derivatives thereof.
    The package offers mean square error optimal bandwidth
    selection and associated point estimators,
    as well as uncertainty quantification based on robust bias correction both
    pointwise
    (e.g., confidence intervals) and uniformly (e.g., confidence bands) over
    evaluation points.
    The methods implemented are boundary adaptive whenever the data is compactly
    supported.
    The package also implements regularized conditional density estimation methods, ensuring the resulting density estimate is non-negative and integrates to one.
    We contrast the functionalities of \texttt{lpcde} with existing open-source
    packages for conditional density estimation,
    and showcase its main features using simulated and real datasets. An abbreviated version of this article is published in \cite{Cattaneo-Chandak-Jansson-Ma_2024_JOSS}.
\end{abstract}

\section{Introduction}

Conditional cumulative distribution functions (CDFs), conditional probability density functions (PDFs), and derivatives thereof, are important
parameters of interest in statistics, econometrics, and other data science
disciplines. This article discusses the main methodological features of the
\proglang{R} package \pkg{lpcde} for estimation of and inference on conditional
CDFs, conditional PDFs, and derivatives thereof,
employing the kernel-based local polynomial smoothing approach introduced in
\citet[CCJM hereafter]{Cattaneo-Chandak-Jansson-Ma_2024_Bernoulli}.

\cite{Wand-Jones_1995_Book},
\cite{Fan-Gijbels_1996_Book},
\cite{simonoff2012smoothing},
and \cite{scott2015multivariate} give textbook introductions to kernel-based
density and local polynomial estimation and inference methods.
The core idea underlying the estimator introduced in CCJM is to use kernel-based
local polynomial smoothing methods to construct an automatically boundary
adaptive estimator for CDFs,
PDFs, and derivatives thereof.
The estimation approach consists of two steps.
The first step estimates the conditional distribution function using standard
local polynomial regression methods,
and the second step applies local polynomial smoothing to the
(non-smooth) local polynomial conditional CDF estimate from the first step to
obtain a smooth estimate of the CDF, PDF, and derivatives thereof.

For the case of PDF estimation, classical estimation approaches typically employ
ratios of unconditional kernel density estimators,
the derivative of kernel-based non-linear distribution function regression
estimators,
or local polynomial estimators based on some preliminary density-like
approximation.
See, for example,
\cite{Fan-Yao-Tong_1996_Biometrika},
\cite{Hall-Wolff-Yao_1999_JASA},
\cite{DeGooijer-Zerom_2003_SN},
\cite{Hall-Racine-Li_2004_JASA}, and references therein.
These approaches are not boundary adaptive unless specific modifications
(e.g., boundary corrected kernels) are introduced.
CCJM's estimator is conceptually different and is boundary adaptive for a
possibly unknown compact support of the data.
Furthermore, the estimator has a simple closed form representation,
which leads to easy and fast implementation.
Unlike some other boundary adaptive procedures,
it does not require pre-processing of data,
and thus avoids the challenges of hyper-parameter tuning:
only one bandwidth parameter needs to be selected for implementation.

Building on the theoretical and methodological work reported in CCJM,
the package \pkg{lpcde} offers data-driven (pointwise and uniform)
estimation and inference methods for conditional CDFs, conditional PDFs,
and derivatives thereof, which are automatically valid at interior,
near-boundary, and boundary points on the support of both the variable of
interest and the conditioning variables. For point estimation, the package
offers mean squared error optimal bandwidth selection and associated
point estimators.
For inference, the package offers valid confidence intervals and confidence
bands based on robust bias-correction techniques
\citep{Calonico-Cattaneo-Farrell_2018_JASA,Calonico-Cattaneo-Farrell_2022_Bernoulli}.
Finally, these statistical procedures can be easily used for visualization and
graphical presentation of smooth empirical CDFs, conditional PDFs, and derivative thereof.
We give an overview of the main methods implemented in the package below,
along with a discussion of more specific implementation issues.
We also showcase the performance of the package with simulated data.

The package \pkg{lpcde} includes two main functions.

\begin{itemize}
    \item \code{lpcde()}:
    This function implements the estimator of interest over a grid of
    evaluation points on the support of the variable of interest and at a
    pre-specified conditioning value.
    The function takes three main inputs:
    data, a bandwidth, and polynomial orders.
    When the bandwidth is not specified by the user,
    the function employs the companion function \code{lpbwcde()} for automatic,
    data-driven bandwidth selection.
    When the polynomial orders are not specified by the user,
    the function employs the next polynomial order relative to the parameter
    of interest.
    For example, for CDF estimation,
    the polynomial orders are set to $\pOrder=\qOrder=1$,
    while for PDF estimation they are set to $\pOrder=2$ and $\qOrder=1$,
    where $\pOrder$ denotes the polynomial order for the variable of interest,
    and $\qOrder$ denotes the polynomial order for the conditioning variables.
    The Epanechnikov kernel is used by default. However, there are
    alternative kernel options that can be provided by the user, if desired.
    This function implements pointwise and uniform inference via robust
    bias-correction methods,
    employing the same grid of points used for point estimation.

    \item \code{lpbwcde()}:
    This function implements pointwise and integrated mean square error (IMSE)
    optimal bandwidth selection for the kernel-based local polynomial smoothing
    methods introduced in CCJM.
    The resulting bandwidth selection procedure leads to an IMSE-rate optimal
    point estimator whenever the difference of polynomial order and derivative
    order of interest is odd (see below for further details).
    This bandwidth choice is also valid,
    and in some cases optimal from a distributional approximation perspective,
    when coupled with robust bias-correction methods for statistical inference.
\end{itemize}

The methods \code{coef()},
\code{confint()},
\code{vcov()},
\code{print()},
\code{plot()}
and \code{summary()}
are supported for objects returned by the \code{lpcde} function,
while the methods \code{coef()}, \code{print()} and \code{summary()}
are supported for objects returned by the \code{lpbwcde} function.
The \code{plot()} function builds on the
\pkg{ggplot2}~\citep{ggplot2} package in \proglang{R} and can be used for illustrations of
conditional CDFs, conditional PDFs or higher order derivatives
and their pointwise or uniform confidence bands
for a given value of the conditioning variable(s).

The package \code{lpcde} contributes to a rather small set of open source
software packages for estimation and inference about
conditional CDF, PDF, and derivatives thereof.
More specifically, we identified two \proglang{R} packages, \pkg{hdrcde}~\citep{hdrcde}, \pkg{haldensify}~\citep{hejazi2022haldensify}, and \pkg{np}~\citep{np}, and one \proglang{Python} package, \pkg{cde}~\citep{rothfuss2019conditional}, which provide related
methodology. There are no open source \proglang{Stata} packages that implement conditional
CDF, PDF, and derivative thereof estimation. Table~\ref{table:rpkgs} summarizes
some of the main differences
between those packages and \pkg{lpcde}.
As is noted in the Table~\ref{table:rpkgs}, \pkg{lpcde} is the only package available across
multiple programming languages
that provides both pointwise and uniform confidence
interval construction that is asymptotically valid, in addition to producing mean square and uniform optimal boundary adaptive point estimates, with the option of ensuring  proper conditional density estimates that are
non-negative and integrate to one. These features are unique contributions of
the package to the \proglang{R} toolkit and, more broadly, the open source
statistical software community.

\medskip
\begin{table}[ht]
\centering
\caption{Comparison of open source software packages for conditional density estimation.}
\resizebox{0.95\columnwidth}{!}{
\begin{tabular}{l|cccccccc}
\hline
\hline
\begin{tabular}[c]{@{}l@{}}Package \\\end{tabular} &
  \begin{tabular}[c]{@{}l@{}} Programming\\language\end{tabular} &
  \begin{tabular}[c]{@{}l@{}} CDF / Derivative\\estimation\end{tabular} &
  \begin{tabular}[c]{@{}l@{}} Regularized\\density\end{tabular} &
  \begin{tabular}[c]{@{}l@{}}Valid at \\ boundary\end{tabular} &
  \begin{tabular}[c]{@{}l@{}}Standard \\ error\end{tabular} &
  \begin{tabular}[c]{@{}l@{}}Valid \\ inference\end{tabular} &
  \begin{tabular}[c]{@{}l@{}}Confidence \\ bands\end{tabular} &
  \begin{tabular}[c]{@{}l@{}}Bandwidth \\ selection\end{tabular} \\
  \hline \hline
  \pkg{\href{https://pkg.robjhyndman.com/hdrcde/}{hdrcde}}\color{black}
&\proglang{R} &$\times$ & $\times$ & $\times$ & $\times$ & $ \times $ & $\times$
& \checkmark \\
  \hline
  \pkg{\href{https://CRAN.R-project.org/package=np}{np}}\color{black}
&\proglang{R} &$\times$ & $\times$ & $\times$ & \checkmark & $\times$ & $\times$
& \checkmark \\
  \hline
  \pkg{\href{https://CRAN.R-project.org/package=haldensify}{haldensify}}\color{black}
&\proglang{R} &$\times$ & $\times$ & $\times$ & \checkmark & $\times$ & $\times$
& \checkmark \\
  \hline
  \pkg{\href{https://github.com/freelunchtheorem/Conditional_Density_Estimation}{cde}}\color{black}
&\proglang{Python} &$\times$ & $\times$ & $\times$ & $\times$ & $\times$ &
$\times$ & \checkmark \\
  \hline \hline
  \pkg{\href{https://cran.r-project.org/web/packages/lpcde/index.html}{lpcde}}\color{black}
&\proglang{R} &\checkmark & \checkmark & \checkmark & \checkmark & \checkmark &
\checkmark & \checkmark \\
  \hline \hline
\end{tabular}%
}
\flushleft
{\small
Notes:
(i) all packages provide conditional PDF point estimation;
(ii) bandwidth selection is done via cross-validation in \pkg{hdrcde} and
\pkg{np},
and using plug-in mean squared error approximations in \pkg{lpcde}.
}
\label{table:rpkgs}
\end{table}

In addition, \cite{Cattaneo-Jansson-Ma_2020_JASA,Cattaneo-Jansson-Ma_2022_JSS,Cattaneo-Jansson-Ma_2024_JOE} develop complementary methods for local polynomial kernel based regression
estimation and inference for \textit{unconditional} densities and higher-order derivatives. These methods and companion statistical software (\pkg{lpdensity}) cannot be used to conduct estimation and inference for \textit{conditional} distributions, densities, and derivative thereof.

The remainder of this article is organized as follows.
Section~\ref{sec:methodology} describes the derivation of our estimator along
with details on how the bandwidth, covariance matrix and confidence intervals
can be constructed. This section also highlights some key computational
considerations when implementing the estimator.
Section~\ref{sec:implementation} discusses how the various functions and
features of the package can be implemented in practice through examples of
code snippets with a toy dataset.
Section~\ref{sec:computation} illustrates the performance of the estimator
through Monte Carlo simulation excercises and compares the performance of
\pkg{lpcde} against the alternative packages identified in
Table~\ref{table:rpkgs} on a real dataset.
Finally, we conclude in Section~\ref{sec:conclusion}.
An abbreviated version of this article is published in \cite{Cattaneo-Chandak-Jansson-Ma_2024_JOSS}, and additional information about the \proglang{R} package \pkg{lpcde},
including replication files and datasets, can be found at
\url{https://nppackages.github.io/lpcde/}.

\section{Methodology}
\label{sec:methodology}

We give an overview of the methodology implemented in \pkg{lpcde};
technical details in full generality can be found in CCJM.
We start by considering a random sample $(Y_1,\bX_1^\Trans),
\dots,(Y_n,\bX_n^\Trans)$ from the continuously distributed random vector
$(Y,\bX^\Trans)\in \Y\times\X$. We assume $\Y\subseteq\mathbb{R}$ is a
$1$-dimensional and
$\X\subseteq\mathbb{R}^{d}$ is a $d$-dimensional possibly,
but not necessarily,
compactly supported set.
The goal is to estimate and conduct inference on the
conditional CDF, PDF, and derivatives thereof, of $Y|\bX$.
Prior to setting up our estimator, the following section establishes all
necessary notation.

\subsection{Notation}
\label{sec:notation}
Notation introduced in this section will be used through the remainder of the
text. Our parameter of interest is
\begin{align*}
    F^{(\mu, \bnu)}(y|\bx)
  = \frac{\partial^{\mu+|\bnu|} }{\partial y^\mu \partial \bx^{\bnu}} F(y|\bx),
  \qquad
  F(y|\bx) = \Prob[Y \leq y | \bX = \bx],
\end{align*}
where $\mu\in\mathbb{N}_0$
denotes the derivative order with respect to the
variable of interest $Y$ and,
employing multi-index notation,
$\bnu\in\mathbb{N}^d_0$ denotes the multi-index for the
corresponding derivatives of interest with respect to the conditioning variables
$\bX$.
For example,
\begin{itemize}
\item $F(y|\bx) = F^{(0, \mathbf{0})}(y|\bx)$ is the conditional CDF of $Y|\bX$;
\item $f(y|\bx) = F^{(1, \mathbf{0})}(y|\bx)$ is the conditional PDF of $Y|\bX$;
\item $f^{(1,0)}(y|\bx) = F^{(2, \mathbf{0})}(y|\bx)$ is the derivative (with respect
  to $y$) of conditional PDF of $Y|\bX$.
\end{itemize}
To simplify the exposition,
we abstract from derivative estimation with respect to the conditioning
variables in $\bX$,
and therefore set $\bnu=\mathbf{0}$ for the rest of this article.
Consequently, we denote
$$F^{(\mu)}(y|\bx)=F^{(\mu, \mathbf{0})}(y|\bx) \ \text{for } \mu\in\mathbb{N}_0.$$
See CCJM for theoretical and methodological results concerning $|\bnu|>0$,
all of which are also implemented in the \proglang{R} package \pkg{lpcde},
thereby allowing for estimation of derivatives with respect to
$\bX$ of the conditional CDF of $Y|\bX$.

The following notation is used in constructing and analysing our estimator:
\begin{itemize}
\item $\be_\ell$ is the conformable $(\ell+1)$-th unit vector.
\item $|\A|$ denotes the cardinality of a set $\A$.
\item $\bq(\bu)$: $\qOrder$-th order polynomial expansion for some
  $\qOrder\in\mathbb{N}$. It is a
  $(\qOrder_d+1)$-dimensional vector collecting the ordered elements
  $\bu^{\bnu}/\bnu!$ for $0\leq|\bnu|\leq \qOrder$, where, employing multi-index
notation, $\bu^{\bnu}=u_1^{\nu_1}u_2^{\nu_2}\cdots u_d^{\nu_d}$,
$\bnu!=\bnu_1!\bnu_2!\cdots \bnu_d!$, $|\bnu|=\nu_1+\nu_2+\dots+\nu_d$,
  and $\qOrder_d = (d+\qOrder)!/(\qOrder!d!) - 1$.
\item $\bp(u)$: $\pOrder$-th order polynomial expansion for some
  $\pOrder\in\mathbb{N}$. It is a $(\pOrder+1)$-dimensional
  vector collecting the ordered elements $u^{\mu}/\mu!$ for $0\leq\mu\leq
  \pOrder$.
\item $K_h(x; u) = K((x-u)/h)/h$, where $K(\cdot)$ is a kernel function and
  $h$ is a bandwidth.
\item $L_h(\bx; \bu)=K_h(x_1-u_1)K_h(x_2-u_2)\cdots K_h(x_d-u_d)$.
\item
  $ \widehat{\bS}_y = \frac{1}{n}\sum_{i=1}^{n}  K_h(y_i; y)
  \bp\Big(\frac{y_i-y}{h}\Big) \bp\Big(\frac{y_i-y}{h}\Big)^\Trans$.
\item
      $ \widehat{\bS}_\bx =
      \frac{1}{n}\sum_{i=1}^{n} L_h(\bx_i; \bx)
      \bq\Big(\frac{\bx_i-\bx}{h}\Big)\bq\Big(\frac{\bx_i-\bx}{h}\Big)^\Trans
$.
\item $  \widehat{\bR}_{y,\bx}
  = \frac{1}{n^2h}\sum_{j=1}^{n}\sum_{i=1}^{n}
  K_h(y_j; y) \bp\Big(\frac{y_j-y}{h}\Big)
  L_h(\bx_i; \bx) \bq\Big(\frac{\bx_i-\bx}{h}\Big)^\Trans
  \Indicator(y_i\leq y_j).
$
\end{itemize}

\subsection{General estimation idea}
\label{sec:general_setup}
The construction of the conditional
CDF, PDF and derivatives thereof
involves two steps.
First,
the conditional distribution function $F(y|\bx)$
is estimated by standard local polynomial methods:
\begin{align}
  \label{eq:step1}
 \widehat{F}_{\qOrder}(y|x) = \be_0^\Trans\widehat{\bgamma}_{\qOrder}(y|\bx), \qquad
 \widehat{\bgamma}_{\qOrder}(y|\bx)
  = \argmin_{\bc\in\mathbb{R}^{\qOrder_d}}
  \sum_{i=1}^n \left(\Indicator(y_i\leq y) - \bq(\bx_i-\bx)^\Trans\bc \right)^2
  L_h(\bx_i-\bx),
\end{align}
Note here that the estimator $\widehat{F}_\qOrder(y|x)$ of $F(y|\bx)$
is not smooth as a function of $y$ and therefore cannot be used to construct an
estimator of the conditional PDF and higher-order derivatives with respect to
$y$.

Therefore, in a second step,
a smoothed (with respect to $y$) estimator of the CDF
and its derivatives is constructed also using local polynomial methods:
for any $0\leq\mu\leq \pOrder$,
\begin{align}
  \label{eq:step2}
  \widehat{F}^{(\mu)}_{\pOrder,\qOrder}(y|\bx)
  &
  = \be_{\mu}^\Trans\widehat{\bbeta}_{\pOrder,\qOrder}(y|\bx),
    \nonumber
  \\
  \widehat{\bbeta}_{\pOrder,\qOrder}(y|\bx)
  &
  = \argmin_{\bb\in\B}
  \sum_{i=1}^n \left(\widehat{F}_\qOrder(y_i|\bx) - \bp(y_i-y)^\Trans\bb \right)^2 K_h(y_i-y),
\end{align}
where $\B$ is some general constraint set.
There are different forms of search space that may be of interest to researchers
based on the application for which the estimator is being used.
For example, it may be necessary that the first
element of $\bb$, corresponding to the conditional PDF estimator, be
nonnegative. In this setting, the set over which $\bb$ is minimized can be
defined as
$\B= \{\bb \in \mathbb{R}^{\pOrder+1}: \be_1^\Trans \bb \geq 0\}$.
This case is studied further in Section~\ref{sec:constrained_estimation}.
On the other hand, if $\B = \mathbb{R}^{\pOrder+1}$, no constraints are imposed
on $\bb$. This is the case that we focus on in the following sections.

\subsection{Point estimation}
\label{sec:pointest_bw}
Solving Equations~\ref{eq:step1} and~\ref{eq:step2} (with $\B=\mathbb{R}^{\pOrder+1}$)
gives a simple closed form for
the general estimator:
\begin{align}
  \label{equation:closed-form expression}
  \widehat{F}^{(\mu)}_{\pOrder,\qOrder}(y|\bx)
  = \be_{\mu}^\Trans\widehat{\bS}_y^{-1} \widehat{\bR}_{y,\bx} \widehat{\bS}_\bx^{-1}\be_{0},
\end{align}
A complete derivation of this closed-form solution is provided in the
supplemental material of CCJM.

In the \proglang{R} package,
for a choice of derivative $\mu$ with respect to $y$
(and a choice of of derivative $\bnu$ with respect of $\bx$,
a choice of polynomial orders $(\pOrder,\qOrder)$,
a choice of bandwidth $h$ and kernel function $K(\cdot)$),
the function \code{lpcde()} implements the estimator
$\widehat{F}^{(\mu)}_{\pOrder,\qOrder}(y|\bx)$
over a grid of points on $\Y$ for a given conditioning evaluation point $\bx$.
By default, the function sets $(\mu,\bnu)=(1,0)$ (conditional PDF),
$\qOrder=1$ (local linear nonsmooth conditional CDF estimation),
$\pOrder=2$ (local quadratic smooth conditional CDF estimation),
and $K(\cdot)$ is to chosen to be the Epanechnikov kernel.
Generally speaking, it is recommended to choose the
local polynomial order such that $\pOrder-\mu$ and $\qOrder-|\bnu|$ are both odd.
Although the second-order Epanechnikov kernel is implemented by default,
the function \code{lpcde()} can also be implemented with
second-order uniform and triangular kernels
by setting the variable \code{kernel\_type} appropriately.
The choice of the kernel does not affect the orders of
the bias and the variance.
Last but not least,
the choice of bandwidth $h$ is important:
by default, whenever $h$ is not supplied by the user,
the function \code{lpcde()} relies on the companion function \code{lpbwcde()},
which implements data-driven bandwidth selection
based on the minimization of the (approximate)
mean squared error of the estimator $\widehat{F}^{(\mu)}_{\pOrder,\qOrder}(y|\bx)$.

In the remainder of this section we review some of the
main statistical properties and inference techniques developed in CCJM
and the computational considerations in implementing these methods
in the package \pkg{lpcde}.

\subsection{Bandwidth selection}
Once we have the closed form of the point estimator,
we can derive the leading bias and variance of the estimator.
The leading bias and variance for odd values of
$\pOrder-\mu$ and $\qOrder-|\bnu|$ take the following form:
\begin{align}
\label{eq:bias}
  &  \text{Bias}\left[\widehat{F}^{(\mu)}(y|\bx)\right]
  = h^{\qOrder+1}\sum_{|\bm|=\qOrder+1}F^{(\mu, \bm)}(y|\bx)B^{(i)}_{\bm}(\bx)
  + h^{\pOrder+1-\mu}F^{(\pOrder+1)}B^{(ii)}_{\pOrder+1}(y),
  \\ &
\label{eq:var}
  \text{Var}\left[\widehat{F}^{(\mu)}(y|\bx)\right]
  = \frac{1}{nh^{d+2\mu+1}} F^{(1)}(y|\bx)V^{(\mu)}_{\pOrder,\qOrder}(y, \bx).
\end{align}
The quantities on the right hand side above implicitly depend on
the kernel function.
It is straightforward to show that both the bias and variance terms
converge in probability to non-random, well-defined limits.
Exact expressions and technical details for other cases can be found in the
supplemental appendix of CCJM.

Equations~\ref{eq:bias} and ~\ref{eq:var} are valid for all evaluation points on the support of the data.
As a result, the pointwise mean squared error (MSE) optimal bandwidth can be approximated as
\begin{align*}
   h_{\pOrder, \qOrder}^{\text{MSE}}(y,\bx)
   = \argmin_{h>0}\left[
   \text{Var}\left[\widehat{F}^{(\mu)}_{\pOrder, \qOrder}(y|\bx)\right]
   + \text{Bias}\left[\widehat{F}^{(\mu)}_{\pOrder, \qOrder}(y|\bx)\right] ^{2}
   \right].
\end{align*}
Under standard regularity conditions, $h^{\text{MSE}}(y,\bx)$ is
MSE-optimal if $\pOrder-\mu$ and $\qOrder-|\bnu|$ are odd.
Precise closed-form expressions for the MSE-optimal bandwidth
can be found in the supplemental appendix of CCJM.
In practice, the MSE-optimal bandwidth is estimated by plugging-in
estimates of the unknown quantities
in Equations~\ref{eq:bias} and ~\ref{eq:var},
given some initial bandwidth choice and then direcetly solving for the optimal
bandwidth.

The IMSE-optimal bandwidth is estimated similarly,
with the main difference being that a set of grid points on the support of $\Y$
is used to approximate the integral.
Detailed expressions are given in the supplemental material of CCJM.
Bandwidth selection is implemented through the \code{lpbwcde()} function.

The number of grid points or specific locations of grid points (default is 19
equally-spaced points over the implied support) can be specified
by the user as an input to both the \code{lpbwcde()} and \code{lpcde()} functions.
For generating quantile-spaced grid points, the flag \code{grid\_spacing} should
be set to \code{`quantile'}.
Users should be aware of possible issues with using equally-spaced grid points
at low-density regions or near boundary points. If a small banwidth is coupled
with grid points that have few data points that can be used for estimation,
the resulting point estimates as well as standard error approximations may have
numerical inaccuracies that cause instability in the output.
We recommend either prior checking of effective sample sizes for the choice of
bandwidth or choosing quantile-spaced grid points.

\subsection{Constrained density estimation}
\label{sec:constrained_estimation}
As mentioned in Section~\ref{sec:general_setup}, some applications may require
that the conditional density estimate satisfy
additional constraints. For example, it may be desirable to ensure that the PDF
estimate be non-negative on the support and integrates to one. Fortunately, our
two-step formulation of the estimator allows for the non-negativity constraint
to be incorporated directly into the second step given in Equation~\ref{eq:step2}:
\begin{align*}
  \widehat{f}_{\mathtt{N}}(y|\bx)
  = \be_{1}^\Trans\widehat{\bbeta}_{\mathtt{N}}(y|\bx), \qquad
  \widehat{\bbeta}_{\mathtt{N}}(y|\bx)
  = \argmin_{\substack{\bb\in\mathbb{R}^{\pOrder+1}:\  \be_{1}^\Trans\bu\geq 0}}
  \sum_{i=1}^n \left(\widehat{F}(y_i|\bx) - \bp(y_i-y)^\Trans\bb \right)^2 K_h(y_i;y),
\end{align*}
where we use the subscript ``$\mathtt{N}$'' to denote the non-negative estimator.
The solution to this modified optimization problem leads to a simple closed form
solution that can be written in terms of the unconstrained estimator
$\widehat{f}(y|\bx)$:
\begin{align*}
  \widehat{f}_{\mathtt{N}}(y|\bx) = \max\big\{ \widehat{f}(y|\bx)\ ,\ 0 \big\}.
\end{align*}
In order to incorporate the constraint that the conditional density estimator also integrates to one,
a global constraint must be imposed on the estimator.
In CCJM, we propose and study a modification
of the $\widehat{f}_{\mathtt{N}}$ based on the Kullback-Leibler divergence,
\begin{align*}
  \widehat{f}_{\mathtt{I}}(y|\bx)
  = \argmin_{g\in\mathcal{G}}
  \text{KL}\big(g \; \big\| \; \widehat{f}_{\mathtt{N}}(\cdot|\bx)\big),
  \qquad
  &\text{where }\text{KL}(g \; \big\| \; f) =
    \int_{\Ysupp} g(y) \log\left(\frac{g(y)}{f(y)}\right) \text{d}y,
\end{align*}
where the subscript ``$\mathtt{I}$'' stands for ``integrating to one''
and
$\mathcal{G} = \{g \geq 0: \int_{\Ysupp} g(y) \text{d}y = 1,\
g(y)=0\text{ for }y\not\in\mathcal{Y}\}$.

Fortunately, $\widehat{f}_{\mathtt{I}}$ can be written in closed form as
\begin{align*}
  \widehat{f}_{\mathtt{I}}(y|\bx)
  =
  \frac{\widehat{f}_{\mathtt{N}}(y|\bx)}
  {\int_{\Ysupp} \widehat{f}_{\mathtt{N}}(u|\bx) \text{d}u}
\end{align*}
Uniform rates of convergence as well as distributional convergence
of both constrained estimators can be established with slight modifications from the
theory that was established for the unconstrained estimator. Crucially, this
means we can construct robust bias-corrected uniform confidence bands for the
constrained estimators as well.
Further details regarding the convergence guarantees of the
constrained estimators are provided in Section 4 of CCJM.

\subsection{Distribution theory and robust bias-corrected inference}
In order to conduct inference,
we first construct a Wald-type test statistic
that has the following distributional convergence
\begin{align*}
    T_{\pOrder, \qOrder}(y,\bx)
  =
  \frac{\widehat{F}^{(\mu)}_{\pOrder, \qOrder}(y|\bx)
  -F^{(\mu)}(y|\bx)}
  {\sqrt{\text{Var}\left[\widehat{F}^{(\mu)}_{\pOrder, \qOrder}(y|\bx)\right]}}
    \rightsquigarrow
    \mathcal{N}(B, 1),
\end{align*}
where $\rightsquigarrow$ denotes weak (distributional) convergence as $h\to0$ and $n\to\infty$,
$\mathcal{N}$ denotes the Gaussian distribution,
and $B$ denotes the standardized asymptotic bias emerging whenever a too
``large'' bandwidth is employed (e.g., when the MSE-optimal or IMSE-optimal
bandwidth is used).
See CCJM for details.

As a result, standard confidence intervals with nominal $(1-\alpha)$ coverage
takes the form:
\begin{align*}
    \text{CI}(y, \bx)
    =
    \left[\widehat{F}^{(\mu)}_{\pOrder, \qOrder}(y|\bx)
    \pm z_{1-\alpha/2}
  \sqrt{\widehat{\text{Var}}\left[\widehat{F}^{(\mu)}_{\pOrder, \qOrder}(y|\bx)\right]}\right],
\end{align*}
where $z_\alpha$ is the $\alpha$-th quantile of the standard normal distribution.
However, for ``large'' bandwidths, this confidence interval would be invalid due
to the asymptotic bias, $B$. In practice, undersmoothing is often used to
address the asymptotic bias present.
However,
\cite{Calonico-Cattaneo-Farrell_2018_JASA,Calonico-Cattaneo-Farrell_2022_Bernoulli}
show that undersmoothing is sub-optimal under the standard assumptions of the model.
Instead, they propose a robust bias-correction (RBC)
technique that has better higher-order approximations
and asymptotically correct coverage probabilities.
RBC requires bias-correction of the point estimator
and then adjusting the variance estimate appropriately to construct a
bias-corrected Wald-type statistic.

For our estimator,
we first correct for the first-order bias by using a point estimator
that is generated by increasing the polynomial order for both variables,
$y$ and $\bx$.
To be specific, we use
$\widehat{F}^{(\mu)}_{\pOrder+1, \qOrder+1}(y|\bx; h^\text{MSE}_{\pOrder, \qOrder})$
in place of
$\widehat{F}^{(\mu)}_{p, q}(y|\bx; h^\text{MSE}_{p, q})$.
The bandwidth used is optimal for the point estimate with
the lower order polynomials.
The asymptotically valid confidence intervals now take the form
\begin{align*}
    \text{CI}^{\text{RBC}}(y, \bx)
    = \left[\widehat{F}^{(\mu)}_{\text{RBC}}(y|\bx)
    \pm z_{1-\alpha/2}
    \sqrt{\widehat{\text{Var}}\left[\widehat{F}^{(\mu)}_{\text{RBC}}(y|\bx) \right]}\right],
\end{align*}
where
$\widehat{F}^{(\mu)}_{\text{RBC}}(y|\bx)
\equiv \widehat{F}^{(\mu)}_{\pOrder+1, \qOrder+1}(y|\bx)
= \widehat{F}^{(\mu)}_{\pOrder, \qOrder}(y|\bx)
-
\widehat{\text{Bias}}\left[\widehat{F}^{(\mu)}_{\pOrder, \qOrder}(y|\bx)\right]$.

Additionally, uniform confidence bands can be constructed as
\begin{align*}
    \text{CB}^{\text{RBC}}(\M) =
    \left\{\left[\widehat{F}^{(\mu)}_{\text{RBC}}(y|\bx)
    \pm  z_{\M, 1-\alpha/2}
  \sqrt{\widehat{\text{Var}}\left[\widehat{F}^{(\mu)}_{\text{RBC}}(y|\bx) \right]}\right],
    \quad y \in \M\right\},
\end{align*}
where
$\M$ is a collection of evaluation points on the support $\Y$ and
$z_{\M, \alpha}$ is the $\alpha$-quantile over the collection of
evaluation points
for a normal distribution centered at 0 and with the same variance-covariance matrix as the estimator.
The critical value $z_{\M, 1-\alpha/2}$, is defined by the upper
$\alpha$ quantile of the supremum of the simulated Gaussian process on the grid
$\M$:
\begin{align*}
z_{\M, \alpha}&=
\inf\left\{ u\geq 0:\ \Prob\left[ \left.
\sup_{y \in \M}
|\widehat{\Z}^{(\mu)}(y|\bx)|\leq u \right|
\text{Data} \right]
\geq 1-\alpha  \right\},\\
\end{align*}
where
$\widehat{Z}^{(\mu)} (y|\bx)
\overset{a}{\thicksim}
\mathcal{N}\Big(0,\widehat{\text{Cov}}\big[\widehat{F}^{(\mu)}_{\text{RBC}}(y|\bx) \big]\Big)$.
The confidence band depends on the entire collection of evaluation points.
In \code{lpcde}, $z_{\M, 1-\alpha/2}$, is estimated by using the
maximum over the grid points as an approximation for the supremum over $\M$.
See CCJM (and its supplemental) for technical details and regularity conditions.

The RBC method leads to confidence intervals/bands that are not
centered at the density point estimates since different order polynomials are
used for the point estimates and for inference.
Thus, it may happen that the point estimates lies outside of the RBC confidence
intervals/bands if the underlying distribution has high curvature at some
evaluation point(s).
One solution in this case is to increase the polynomial orders $\pOrder$ and
$\qOrder$, or to use a smaller-than-optimal bandwidth.

\subsection{Implementation of the covariance estimator}
\label{sec:cov_theory}
Implementing the variance estimator for both estimating the MSE-optimal
bandwidth and constructing confidence intervals, requires careful consideration.
It is particularly crucial to consider the computational cost of estimating the
variance when employing uniform
confidence bands, which requires the construction of the full
$|\M|\times |\M|$ covariance matrix in order to approximate
the critical value $z_{\M, 1-\alpha/2}$.
The discussion in this section focuses only on the covariance matrix estimation
for the conditional PDF, $\widehat{f}(y|\bx)$, purely for simplicity of presentation.

The standard plug-in estimator (constructed by estimating the unknown quantities
in Equation~\ref{eq:var}) that is proposed and studied in CCJM has a
computational complexity of $O(|\M|^2n^4)$.
For large datasets and a fine grid of evaluation points, this is
prohibitively slow to run in practice.
As a result the default covariance estimator used in \code{lpcde} implements a
significantly faster jackknife covariance estimatior.
The construction of the jackknife estimator relies on the fact that the
closed-form of the estimator $\widehat{f}$ can be written as
a V-statistic to which the Hoeffding decomposition can be applied.
The covariance expression then decomposes to a sum of two independent functions
that depend on the evaluation points and a small subset of the data in the
neighborhood of the evaluation points, thus reducing the computational cost to
only $O(|\M|^2(nh)^2)$.

Here we provide a sketch of how this estimator is constructed.
The interested reader can find a complete derivation in
Section 6 of the Supplementary Material of CCJM.

We start by first observing that the estimator
$\hat{f}(y|\bx)$
is a V-statistic:
\begin{align}
  \nonumber
\hat{f}(y|\bx)
  &= \frac{1}{n^2h} \sum_{i,j}
    \Indicator(y_i\leq y_j)
    \be^{\Trans}_{1} \hat\bS_{y}^\inv \bp\left( \frac{y_j-y}{h} \right)
    K_h\left( y_j; y \right)
    \bq^{\Trans}\left( \frac{\bx_i-\bx}{h} \right)
    L_h\left(\bx_i;\bx\right)
    \hat\bS_{\bx}^\inv
   \be_{0}
  \\&
  =
  \label{eq:vstat}
  \frac{1}{n^2} \sumIN
  a(y_{i}, y) b(\bx_{i}, \bx)
  +
  \frac{1}{n^2} \sum_{1\leq i \neq j \leq n}
  \Indicator(y_i\leq y_j)
  a(y_{j}, y) b(\bx_{i}, \bx) ,
\end{align}
where
\begin{align*}
  a(y_{i}, y)
  &=
    h^{-2}\be^{\Trans}_{1} \hat\bS_{y}^\inv \bp \left(
    \frac{y_i-y}{h} \right) K\left( \frac{y_i-y}{h} \right) ,
  \\
  b(\bx_{i}, \bx)
  &=
    h^{-d} \be^{\Trans}_{0}\hat\bS_{\bx}^\inv \bQ \left(
    \frac{\bx_i-\bx}{h} \right) L\left( \frac{\bx_i-\bx}{h} \right) .
\end{align*}
The scalar functions $a(\cdot)$ and $b(\cdot)$ are are non-zero only
for data points that are within $h$ distance of the evaluation point, a feature that the package \pkg{lpcde} leverages explicitly to improve numerical performance in applications.
The second term in Equation~\ref{eq:vstat} can now be symmetrized and treated as a
U-statistic.
Then, the Hoeffding decomposition can be applied and plugged back into
Equation~\ref{eq:vstat}.
This leads to a natural alternative jackknife covariance estimator, which is
simple to write and computationally efficient:
\begin{align*}
  \hat{\mathsf{C}}(\eval, y', \bx')
  = \frac{1}{n} \sumIN
  \hat{L}_{(i)}(\eval)
  \hat{L}_{(i)}(y', \bx').
\end{align*}
where
\begin{align*}
  \hat{L}_{(i)}(\eval)
  =  \frac{2}{n-1} \sum_{j \neq i} \left(u_{i,j}
  - \hat{f}(y|\bx) \right).
\end{align*}
and $u_{i,j} =\frac{1}{2} ( \Indicator ( y_i \leq y_j) a(y_{j}, y)b(\bx_{i}, \bx)
  + \Indicator(y_j \leq y_i) a(y_{i}, y)b(\bx_{i}, \bx) )/2$. In particular, if the two evaluation points are equivalent, then
we return the (approximately jackknife) variance estimator
\begin{align*}
  \hat{\mathsf{V}}(\eval)
  \equiv
  \hat{\mathsf{C}}(\eval, \eval)
  = \frac{1}{n-1} \sumIN \hat{L}^{2}_{(i)}(\eval).
\end{align*}
It can be easily verified that this jackknife covariance estimator is
asymptotically equivalent to the theoretical variance expression in Equation~\ref{eq:var}.

\section{Implementation}\label{sec:implementation}
In this section we discuss how each of the functions provided
in \pkg{lpcde} can be used with the aid of code snippets on a simulated dataset.
We consider a bi-variate jointly normal
data generating process with mean $0$ and variance $1$.

\subsection{Density estimation}
The function \code{lpcde()} provides information on
point estimates,
standard errors
and confidence interval or bands
for a given value of
$\bx$ over a range of grid points for $y$.
If the grid points are not provided by the user,
the function chooses nineteen equally-spaced grid points over the
implied support of the data
and, if no bandwidth is provided, computes the rule-of-thumb MSE bandwidth at each point.

The following example estimates the conditional density at $\bx = 0$,
with a fixed bandwidth of $1$,
using the default local polynomial approximation $\pOrder = 2,\ \qOrder = 1$.
RBC confidence intervals over the grid are also computed,
in this case using the default polynomial orders $\pOrder = 3,\ \qOrder = 2$.
\begin{CodeInput}
  R> set.seed(42)
  R> n = 1000
  R> x_data = as.matrix(stats::rnorm(n, mean = 0, sd = 1))
  R> y_data = as.matrix(stats::rnorm(n, mean = x_data, sd = 1))
  R> y_grid = seq(from = -2, to = 2, length.out = 10)
  R> model1 = lpcde(y_data = y_data, x_data = x_data, y_grid = y_grid, x = 0,
  +    bw = 1, rbc = TRUE)
  R> summary(model1)
\end{CodeInput}

The function returns an object of type \code{lpcde}.
Standard \proglang{R} methods,
\code{coef()},
\code{confint()},
\code{vcov()},
\code{print()},
\code{plot()}
and \code{summary()},
can be used on objects of type \code{lpcde} to understand the output.

Below we reproduce the output of running the \code{summary} command on
\code{model1}.
The first part of the summary output provides
basic information about some of the options specified to the function.
The second part provides relevant information
for each point estimate generated in a table with $7$ columns,
(i) grid evaluation points,
(ii) bandwidth used at each point,
(iii) effective number of data points used to generate the point estimate,
(iv) point estimate,
(v) standard error,
(vi) lower $(1-\alpha)$-confidence interval,
and,
(vii) upper $(1-\alpha)$-confidence interval.
\begin{CodeOutput}
Call: lpcde

Sample size                                           1000
Polynomial order for Y point estimation      (p=)     2
Polynomial order for X point estimation      (q=)     1
Density function estimated                   (mu=)    1
Order of derivative estimated for covariates (nu=)    0
Kernel function                                       epanechnikov
Bandwidth method

============================================================================
                                     Point      Std.       Robust B.C.
Index     Grid      B.W.   Eff.n      Est.     Error      [ 95\% C.I. ]
============================================================================
1      -2.0000    1.0000     132    0.0768    0.0126     0.0043 ,  0.1108
2      -1.5556    1.0000     211    0.1446    0.0089     0.0834 ,  0.1693
3      -1.1111    1.0000     304    0.2255    0.0064     0.2135 ,  0.2829
4      -0.6667    1.0000     370    0.2982    0.0051     0.2964 ,  0.3584
5      -0.2222    1.0000     411    0.3407    0.0047     0.3410 ,  0.3981
----------------------------------------------------------------------------
6       0.2222    1.0000     409    0.3397    0.0044     0.3576 ,  0.4113
7       0.6667    1.0000     359    0.2958    0.0050     0.2874 ,  0.3447
8       1.1111    1.0000     279    0.2229    0.0064     0.1916 ,  0.2574
9       1.5556    1.0000     186    0.1390    0.0082     0.0818 ,  0.1656
10      2.0000    1.0000     117    0.0663    0.0112    -0.0291 ,  0.0886
----------------------------------------------------------------------------
============================================================================
\end{CodeOutput}
By default, the function provides estimates according to the original
formulation of the estimator $\widehat{f}(y|\bx)$.
If a constrained density estimate that is non-negative and integrates to one
($\widehat{f}_{\mathtt{I}}$ as defined in
Section~\ref{sec:constrained_estimation}) is desired, the flags \code{nonneg}
and \code{normalize} can be turned on.
\begin{CodeInput}
    R> model_reg = lpcde(y_data = y_data, x_data = x_data, y_grid = y_grid,
    +     x = 0, bw = 1, nonneg = TRUE, normalize = TRUE)
\end{CodeInput}
Figure~\ref{fig:norm_plt} shows how the estimates differ when the
additional constraints are imposed:
\vspace{-0.3cm}
\begin{figure}[H]
  \centering
  \includegraphics[width=0.5\textwidth]{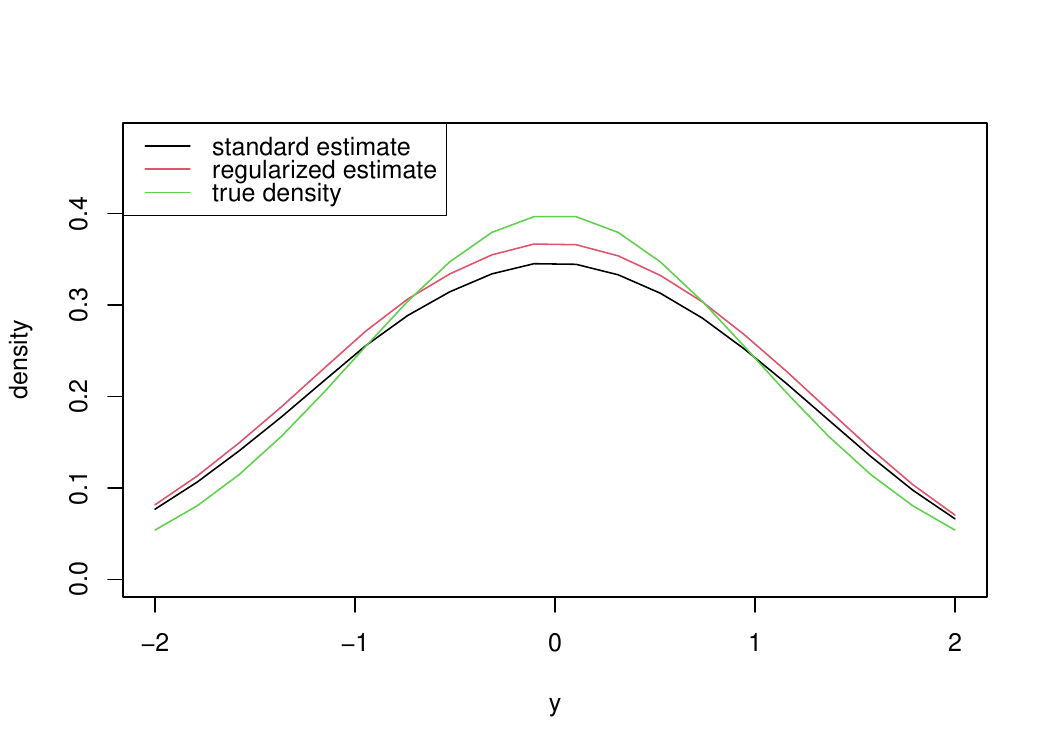}
  \caption{Comparing standard density estimate with normalized estimate.}
  \label{fig:norm_plt}
\end{figure}

\subsection{Out-of-sample prediction}
The \texttt{lpcde} function can be directly used for prediction on a new dataset.
Below is a simple illustration of how a researcher may want to implement this.

Suppose we want to use some data to \textit{train} the estimator and then we
would like to \textit{test} it on an unseen dataset.
In this case, a common method is to randomly subset the data
into training and testing.
We assume the same simulation set up as in the previous section.
The data is split with $95\%$ for training and $5\%$ for testing
\begin{CodeInput}
   R> sample = sample(c(TRUE, FALSE), nrow(y_data), replace=TRUE, prob=c(0.95,0.05))
   R> y_train  = y_data[sample, ]
   R> x_train = x_data[sample, ]
   R> y_test   = y_data[!sample, ]
\end{CodeInput}
Now the \texttt{y\_test} sample can be used directly as the grid of evaluation
points for \texttt{lpcde}:
\begin{CodeInput}
   R> prediction_model = lpcde::lpcde(x_data=x_train, y_data=y_train,
   +    y_grid=y_test, x=0.5, bw=0.5, cov_flag="off")
\end{CodeInput}

\subsection{Covariance estimation}
As noted in Section~\ref{sec:cov_theory}, estimating the full covariance matrix
can be computationally intensive. In order to allow full flexibility in
application of this functionality, an optional input \texttt{cov\_flag}
to the \texttt{lpcde} function can be used. This input can take on three
different values:

\begin{enumerate}[label=(\alph*)]
\item \texttt{"full"}: the function will compute the entire covariance matrix
  (and therefore allow confidence interval and band construction),
\item \texttt{"diag"}: this will only compute the diagonal entries of the
  covariance matrix (i.e. the standard errors, only pointwise confidence
  intervals can be computed), and
\item \texttt{"off"}: no entries of the covariance matrix are estimated.
  Inference tools will be unavailable.
\end{enumerate}

\subsection{Plotting}
\begin{figure}[H]
  \centering
  \includegraphics[width=0.5\textwidth]{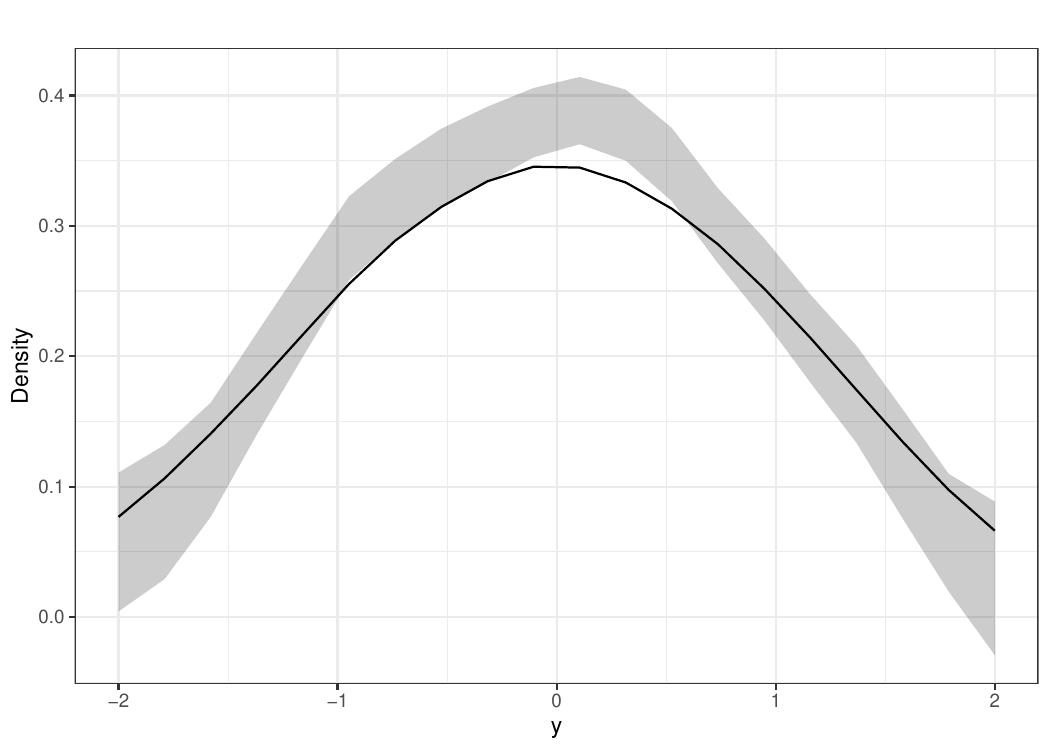}
  \caption{A simple density plot with robust 95\% confidence bands.}
  \label{fig:sample_plot}
\end{figure}

The \code{plot()} function uses the \pkg{ggplot2} package
with objects of type \code{lpcde} to produce illustrations of
point estimates and confidence intervals and/or bands.
A simple plot of the conditional PDF with
95\% confidence intervals
can be generated by running the following code.
\begin{CodeInput}
  R> plot(model1, CIuniform = TRUE, rbc = TRUE, xlabel = "y")
\end{CodeInput}
This code snippet produces an image of the type shown in Figure~\ref{fig:sample_plot}.

By default the \code{plot()} function plots pointwise confidence intervals at
$95\%$ level with the point estimates.
Additional options for confidence levels,
bands and RBC inference are detailed in the package manual.
Editing other visual aspects of the plots
can be done by providing standard inputs to \pkg{ggplot2} functions.

\subsection{Bandwidth selection}
\code{lpbwcde()} implements
the rule-of-thumb MSE- and IMSE- bandwidth selection
by implementing the formulae provided in Section~\ref{sec:pointest_bw}.

By default \code{lpbwcde()} computes the
rule-of-thumb MSE optimal bandwidth for the conditional PDF with
locally quadratic polynomial in $y$ and
locally linear polynomial in $\bx$ and
Epanechnikov kernel on
nineteen equally-spaced grid points on the implied support of $\Y$
determined by the observed data.
The output of this function is similar to that of
\code{lpcde()} and provides basic information for the data and options specified.
The summary of objects returned by this function
additionally provides a table with three columns:
(i) \code{y\_grid}:
values of the grid points for which the bandwidth is estimated,
(ii) \code{B.W.}:
the estimated bandwidth corresponding to each grid point, and
(iii) \code{Eff.n.}:
the number of effective data points at each evaluation point given the
estimated bandwidth.
An example of standard bandwidth selection
is provided in the following output.
\begin{CodeInput}
    R> model2 = lpbwcde(y_data = y_data, x_data = x_data, x = 0,
    +    y_grid = y_grid)
    R> summary(model2)
\end{CodeInput}
\vspace{-0.1cm}
\begin{CodeOutput}
  Call: lpbwcde

  Sample size                                           1000
  Polynomial order for Y point estimation      (p=)     2
  Polynomial order for X point estimation      (q=)     1
  Density function estimated                   (mu=)    1
  Order of derivative estimated for covariates (nu=)    0
  Kernel function                                       epanechnikov
  Bandwidth method                                      mse-rot

  ==================================
  Index     y_grid      B.W.   Eff.n
  ==================================
  1      -2.0000    1.0250      76
  2      -1.5556    1.1594     238
  3      -1.1111    2.0298     808
  4      -0.6667    1.2968     615
  5      -0.2222    1.0609     560
  ----------------------------------
  6       0.2222    1.0634     558
  7       0.6667    1.3103     607
  8       1.1111    1.9603     774
  9       1.5556    1.1566     219
  10      2.0000    1.0274      71
  ----------------------------------
  ==================================
\end{CodeOutput}
The estimated bandwidth from this function can be used as bandwidth input to
\code{lpcde()} directly by using the option of
\code{bwselect} to specify bandwidth selection type instead of running
\code{lpbwcde()} first.

\section{Computational performance}
\label{sec:computation}
In this section we demonstrate the performance of the \pkg{lpcde} package.
We start with a simulated dataset analysis to showcase each of the inference
features. Then, we compare the performance of our package with the existing
conditional density estimators in \proglang{R} (as identified in
Table~\ref{table:rpkgs}) on the Iris dataset.

\subsection{Simulations}
\label{sec:simulations}
In this section we illustrate the effectiveness of our estimator with a Monte
Carlo study.
For the sake of simplicity, we set $d=1$ and assume that $\bx$ and $y$ are
simulated by a joint normal distribution truncated on $[-1.5, 1.5]^2$.
We simulate $100$ data sets of
$2000$ independent samples each.
The point estimates are generated at three distinct values that are
characterized by their location
(a) interior ($0$), (b) near-boundary ($0.8$), and (c) at-boundary ($1.5$)
relative to the implied boundary of the data.

For each conditional value, we present the average bandwidth, average bias,
standard deviation, $95\%$ coverage,
and width of the confidence intervals across the simulated datasets.
We present these results for both the
standard estimate (rows ``WBC'') which is generated with a
quadratic polynomial ($\pOrder=2$) with respect to the variable $y$,
and linear polynomial ($\qOrder=1$) with respect to the variable $\bx$,
as well as the robust bias-corrected estimates (rows ``RBC'')
which uses cubic polynomial ($\pOrder=3$) for $y$
and quadratic polynomial ($\qOrder=2$) for $\bx$.

Table~\ref{table:pw_table} presents the
results of this simulated study.
The first four columns of the table present average
pointwise MSE-optimal bandwidth used in estimation ($\widehat{h}_{\text{MSE}}$),
bias,
standard error (SE) and root mean squared-error (RMSE).
The last four columns are the
average pointwise confidence interval coverage
and width (AW) of the confidence interval
for the standard estimate and inference method (``WBC'')
and robust bias-corrected estimate and inference (``RBC'').
\begin{table}[H]
\centering
\resizebox{0.75\columnwidth}{!}{%
\begin{tabular}{c|cccc|cc|cc}
\hline \hline
& & & & & \multicolumn{2}{c|} { Coverage } & \multicolumn{2}{c} { AW} \\
\hline
Eval. point & $\widehat{h}_{\text{MSE}}$ & Bias & SE & RMSE & WBC & RBC & WBC & RBC \\
\hline
\multicolumn{9}{c}{$\bx=0$}\\
\hline
  $y=0$ & 0.48 & 0.01 & 0.02 & 0.03 & 70 & 92 & 0.07 & 0.22 \\
  $y=0.8$ & 0.55 & 0.01 & 0.01 & 0.02 & 79 & 94 & 0.05 & 0.17 \\
  $y=1.5$ & 0.78 & 0.01 & 0.01 & 0.02 & 56 & 95 & 0.03 & 0.08 \\
\hline
\multicolumn{9}{c}{$\bx=0.8$}\\
\hline
  $y=0$ & 0.65 & 0.01 & 0.01 & 0.02 & 78 & 94 & 0.05 & 0.16 \\
  $y=0.8$ & 0.60 & 0.02 & 0.01 & 0.03 & 53 & 96 & 0.06 & 0.19 \\
  $y=1.5$ & 0.68 & 0.01 & 0.01 & 0.02 & 75 & 94 & 0.05 & 0.15 \\
\hline
\multicolumn{9}{c}{$\bx=1.5$}\\
\hline
  $y=0$ & 1.00 & 0.02 & 0.01 & 0.02 & 49 & 92 & 0.04 & 0.12 \\
  $y=0.8$ & 0.90 & 0.01 & 0.01 & 0.03 & 73 & 93 & 0.05 & 0.15 \\
  $y=1.5$ & 0.90 & 0.04 & 0.02 & 0.05 & 29 & 95 & 0.06 & 0.17 \\
\hline
\hline
\end{tabular}%
}
\caption{Pointwise results
\\
WBC: without bias-correction, RBC: robust bias-corrected. }
\label{table:pw_table}
\end{table}
Note that robust bias-corrected inference produces accurate empirical coverage
across all pointwise combinations. As such, we recommend users employ robust
bias-corrected estimates for improved reliability of results.

Next, we test the bandwidth selection
by simulating point estimation and coverage at varying
bandwidth values.
We choose the range of bandwidth values to be
between $0.5$ and $1.3$ times the average MSE-optimal bandwidth ($\widehat{h}_{\text{MSE}}$).
Table~\ref{table:bw_sel} presents the average
bias,
standard error (SE),
root mean-squared error (RMSE),
pointwise coverage rate (CR)
and
average width of confidence intervals (AW)
for 100 simulations
at the point
$y=0,\ \bx=0$.
\begin{table}[H]
\centering
\begin{tabular}{c|c|ccc|cc|cc}
  \hline
  $\times \widehat{h}_{\text{MSE}}$ & $\widehat{h}$ & Bias & SE & RMSE & WBC CR & RBC CR & WBC AW & RBC AW \\
  \hline
  0.5 & 0.24 & 0.00 & 0.13 & 0.15 & 100.00 & 100.00 & 0.53 & 1.76 \\
  0.6 & 0.29 & 0.01 & 0.08 & 0.09 & 100.00 & 100.00 & 0.30 & 1.01 \\
  0.7 & 0.34 & 0.01 & 0.05 & 0.06 & 97.00 & 100.00 & 0.19 & 0.64 \\
  0.8 & 0.38 & 0.01 & 0.03 & 0.05 & 89.00 & 100.00 & 0.13 & 0.43 \\
  0.9 & 0.43 & 0.01 & 0.02 & 0.04 & 80.00 & 98.00 & 0.09 & 0.30 \\
  1 & 0.48 & 0.01 & 0.02 & 0.03 & 70.00 & 92.00 & 0.07 & 0.22 \\
  1.1 & 0.53 & 0.02 & 0.01 & 0.03 & 57.00 & 85.00 & 0.05 & 0.17 \\
  1.2 & 0.58 & 0.02 & 0.01 & 0.03 & 40.00 & 76.00 & 0.04 & 0.13 \\
  1.3 & 0.62 & 0.02 & 0.01 & 0.03 & 28.00 & 72.00 & 0.03 & 0.10 \\
  1.4 & 0.67 & 0.03 & 0.01 & 0.03 & 17.00 & 68.00 & 0.03 & 0.08 \\
  1.5 & 0.72 & 0.03 & 0.01 & 0.03 & 7.00 & 64.00 & 0.02 & 0.07 \\
  \hline
\end{tabular}
\caption{Bandwidth selection at interior point ($y=0, \bx=0$).
\\
WBC: without bias-correction, RBC: robust bias-corrected. }
\label{table:bw_sel}
\end{table}

\subsection{Comparative analysis}\label{sec:comparison}
\begin{figure}[H]
  \centering
  \includegraphics[width=0.6\textwidth]{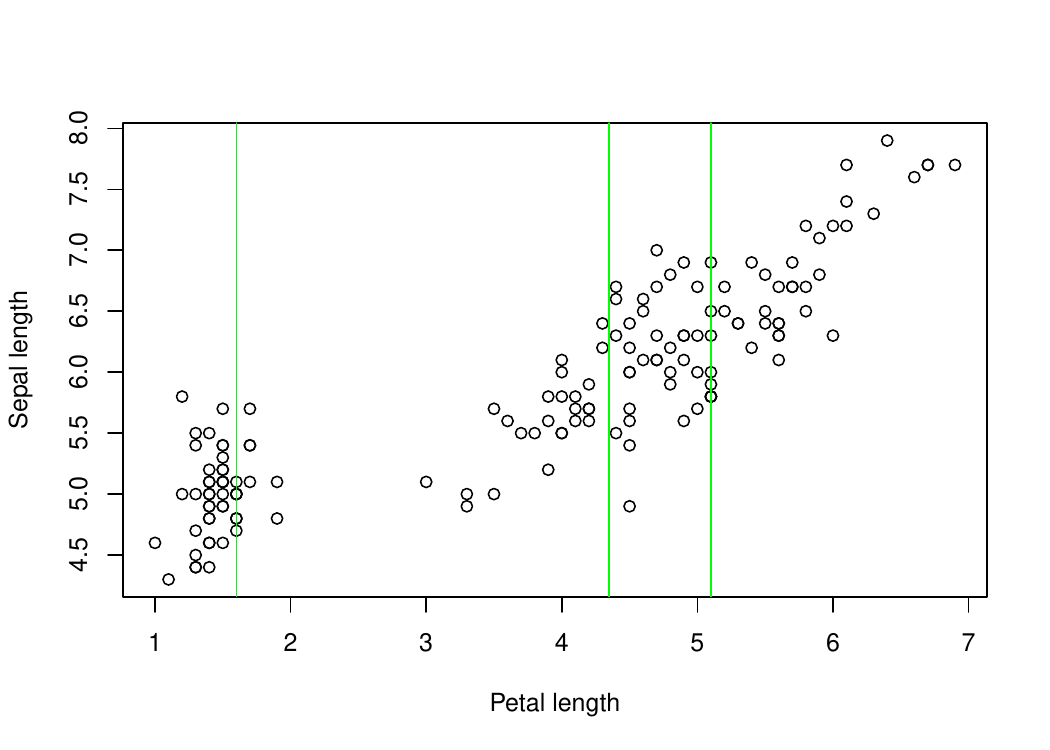}
  \caption{Scatter plot with conditioning values.}
  \label{fig:scatter}
\end{figure}

We now turn to comparing the performance of \code{lpcde} against the other
open source \proglang{R} packages available at the time of writing this article.
To compare the performance of these packages, we consider the Iris dataset which
is available as a default dataset in \proglang{R}.
For this study, we estimate the distribution of the \textit{Sepal length}
feature conditional on the \textit{Petal length} feature.
The conditioning values are chosen based on the $25$-th ($1.6$), $50$-th ($4.35$)
and $75$-th ($5.1$) quantiles of the \textit{Petal length}.
Figure~\ref{fig:scatter} shows a scatter plot of the data with vertical lines
denoting the conditional values at which the \textit{Sepal length} density will
be estimated.

\begin{figure}[H]
  \begin{subfigure}[b]{0.5\linewidth}
    \centering
    \includegraphics[width=0.8\linewidth]{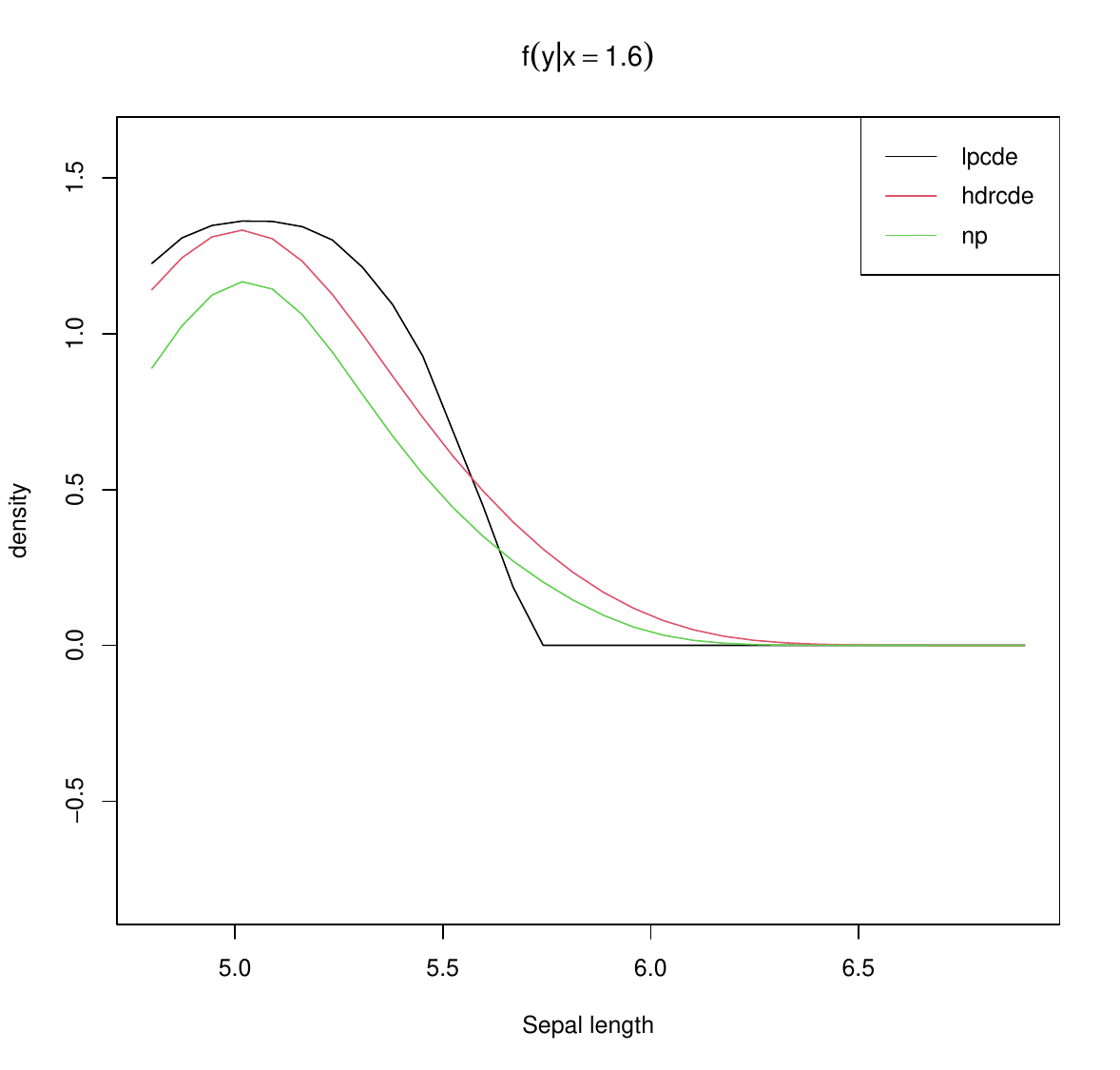}
    \caption{Conditional density estimates.}
    \label{fig:q1_cde}
  \end{subfigure}
  \begin{subfigure}[b]{0.5\linewidth}
    \centering
    \includegraphics[width=0.7\linewidth]{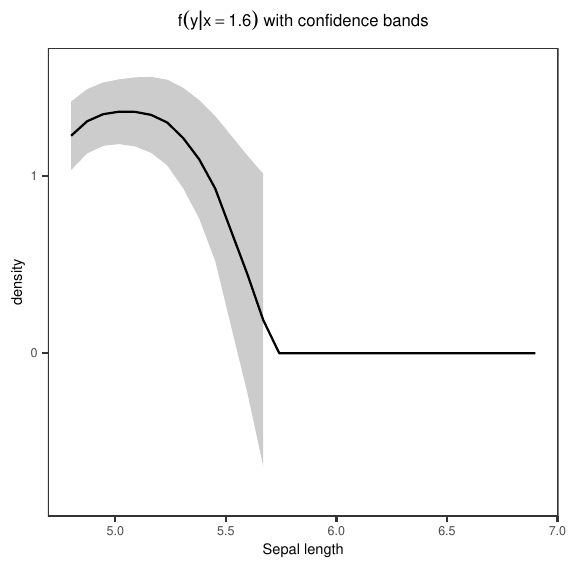}
    \vspace{0.3cm}
    \caption{\code{lpcde} estimate with uniform confidence bands.}
    \label{fig:q1_cb}
  \end{subfigure}
  \caption{Conditional density estimates from each implementation conditioning
    at \textit{Petal length} $= 1.6$.}
  \label{fig:q1}
\end{figure}
From Figure~\ref{fig:scatter}, it is clear that the conditional expectation of
the \textit{Sepal length} shifts across the three evaluation points.
Furthermore, at the conditional value of $1.6$, there are very few data points
in a reasonable neighbourhood that can be used to construct the estimates. We
expect this to affect the standard error and resulting confidence intervals.

\begin{figure}[H]
  \begin{subfigure}[b]{0.5\linewidth}
    \centering
    \includegraphics[width=0.8\linewidth]{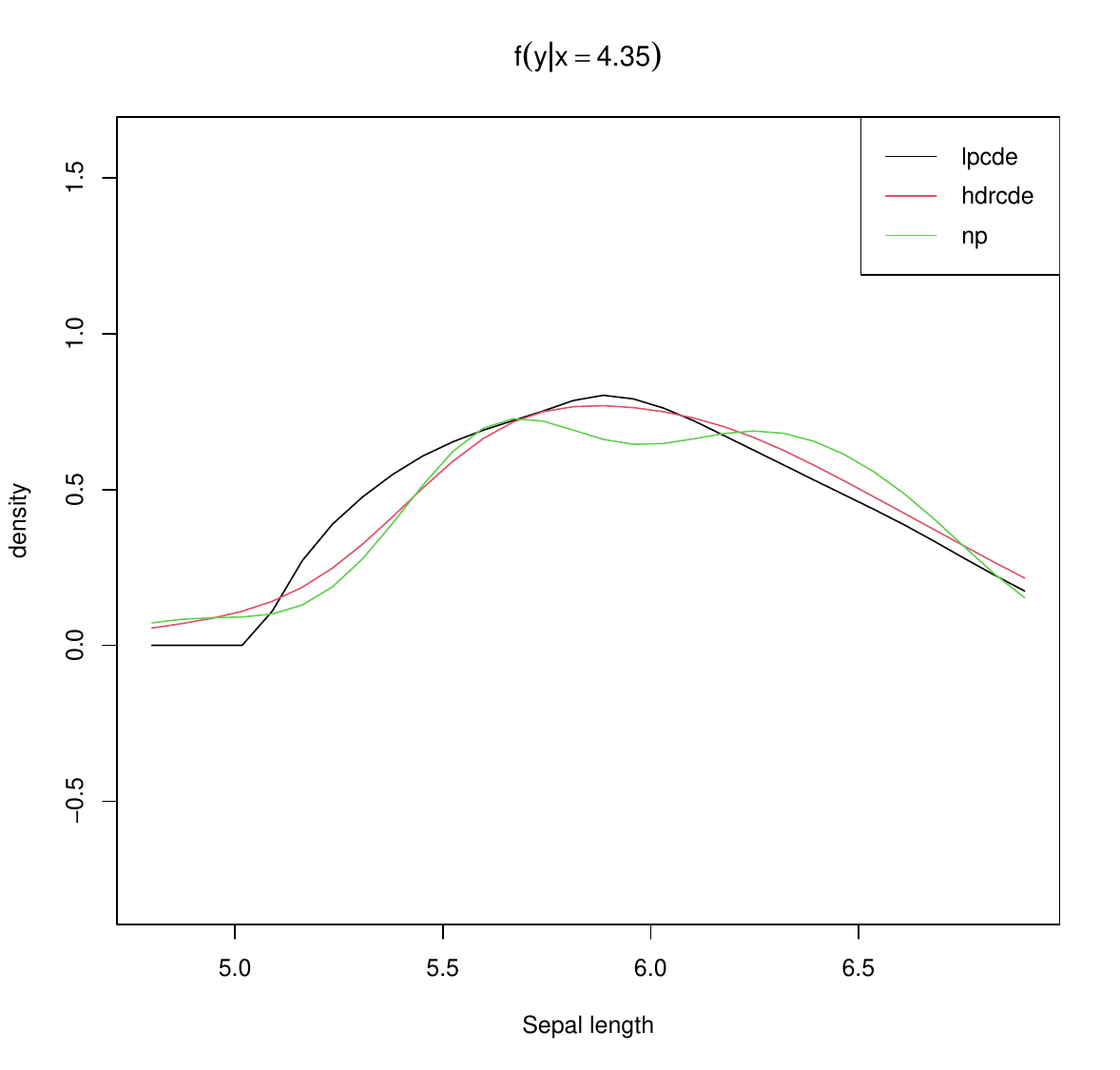}
    \caption{Conditional density estimates.}
    \label{fig:q2_cde}
  \end{subfigure}
  \begin{subfigure}[b]{0.5\linewidth}
    \centering
    \includegraphics[width=0.7\linewidth]{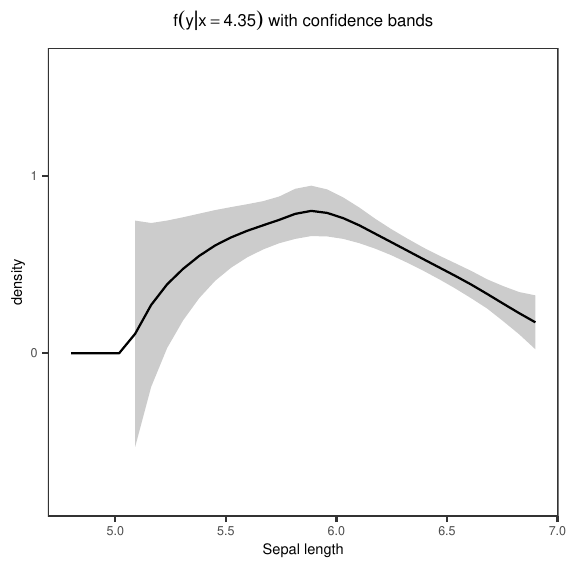}
    \vspace{0.3cm}
    \caption{\code{lpcde} estimate with uniform confidence bands.}
    \label{fig:q2_cb}
  \end{subfigure}
  \caption{Conditional density estimates from each implementation conditioning
    at \textit{Petal length} $ =4.35$.}
  \label{fig:q2}
\end{figure}

\begin{figure}[H]
  \begin{subfigure}[b]{0.5\linewidth}
    \centering
    \includegraphics[width=0.8\linewidth]{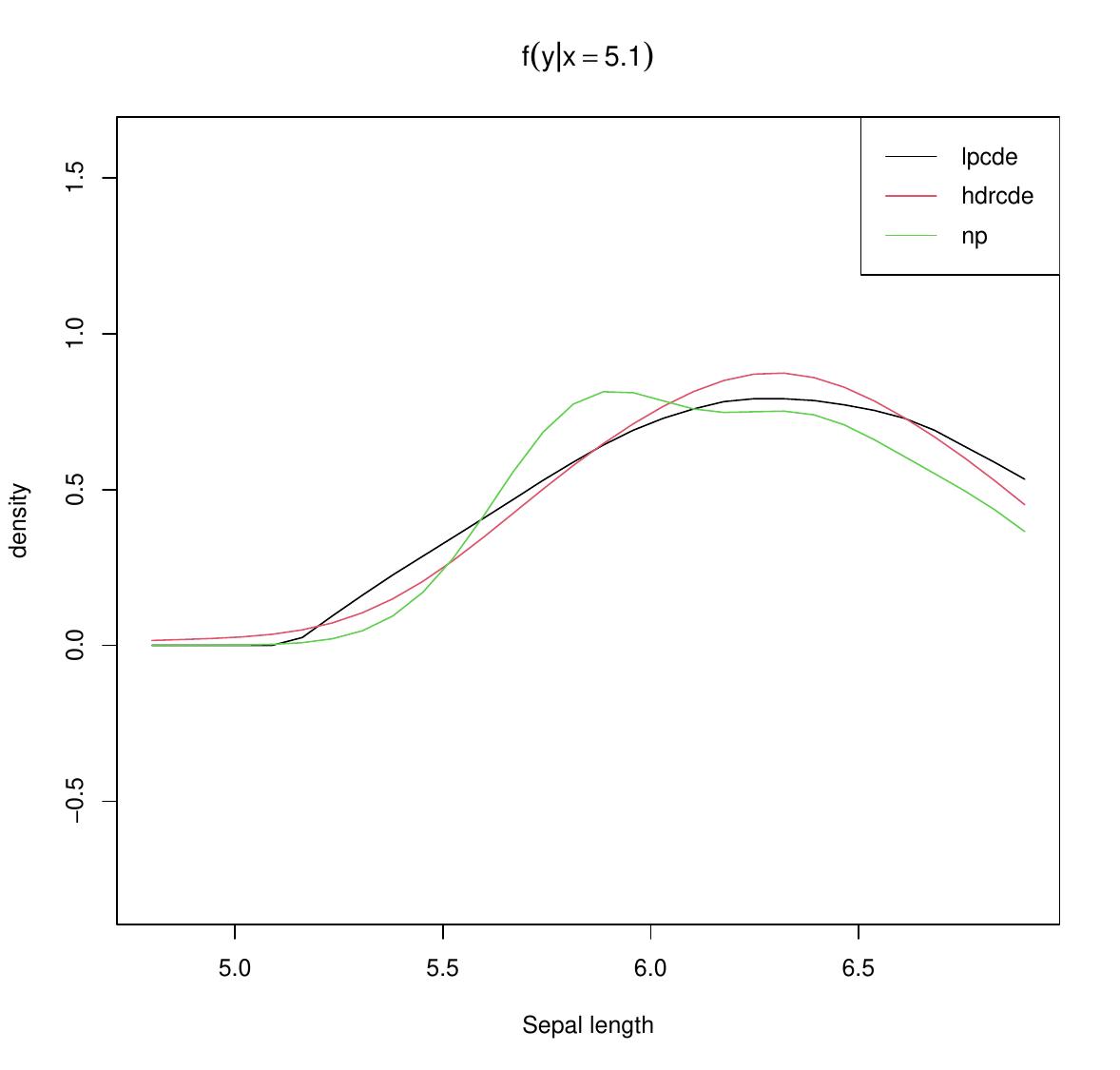}
    \caption{Conditional density estimates.}
    \label{fig:q3_cde}
  \end{subfigure}
  \begin{subfigure}[b]{0.5\linewidth}
    \centering
    \includegraphics[width=0.7\linewidth]{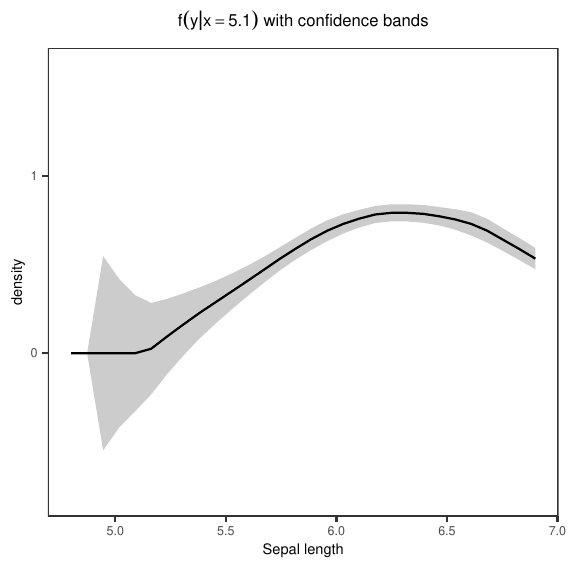}
    \vspace{0.3cm}
    \caption{\code{lpcde} estimate with uniform confidence bands.}
    \label{fig:q3_cb}
  \end{subfigure}
  \caption{Conditional density estimates from each implementation conditioning
    at \textit{Petal length} $= 5.1$.}
  \label{fig:q3}
\end{figure}

We now plot the conditional distributions of each of the three estimators.
Since \pkg{lpcde} is the only package that provides
confidence interval construction, we additionally plot the \code{lpcde} estimate
with the pointwise confidence intervals. Note that the confidence
intervals are only constructed for estimates that are generated with more that
$15$ data points as we believe standard errors on estimates generated with fewer
data points will be unreliable.
Figures~\ref{fig:q1},~\ref{fig:q2} and ~\ref{fig:q3} show the estimated densities
for condtioning at $1.6$, $4.35$ and $5.1$, respectively.
Figures~\ref{fig:q1_cde},~\ref{fig:q2_cde}, ~\ref{fig:q3_cde} compare directly
the estimates generated by the default implementations from each of the three
packages (\pkg{hdrcde}, \pkg{np} and \pkg{lpcde}).
Figures~\ref{fig:q1_cb}, ~\ref{fig:q2_cb}, ~\ref{fig:q3_cb} illustrate the
\pkg{lpcde} estimate with confidence intervals using the default plotting
implementation provided in the package (pointwise, non bias-corrected confidence
intervals).

From the plots in Figures~\ref{fig:q1_cde},~\ref{fig:q2_cde}, ~\ref{fig:q3_cde},
the three estimators largely present the same expected trends of the density
function. One observation that may be of interest is the that \pkg{np} estimate
for $x=4.35$ and $x=5.1$ is slightly bi-modal, which is not reflected in the
other two estimators and arguably is not present in the raw data (see
Figure~\ref{fig:scatter}). Furthermore, the \pkg{np} estimator does not produce a
valid density estimate in that the estimator does not integrate to 1 for any of
conditioning values. On the other hand, \pkg{hdrcde} produces valid density
estimates and is very similar to the estimates of \pkg{lpcde}.
Given that \pkg{hdrcde} does not provide inference tools, we cannot compare the
two packages further.

\section{Conclusion}\label{sec:conclusion}
This article introduced the software package
\pkg{lpcde}, which implements local polynomial kernel based regression
estimation and inference for conditional densities and higher-order derivatives.
This package is currently
the only open source estimator that provides adaptive conditional density
estimation with robust bias-correction and pointwise confidence interval and
uniform confidence band construction, providing users with tools to better
understand the reliability of their analysis.
See \cite{Cattaneo-Chandak-Jansson-Ma_2024_JOSS} for an abbreviated published
version of this article. Additional information and replication files can be
found at \url{https://nppackages.github.io/lpcde/}.

\section*{Acknowledgments}
The authors thank the reviewers of the \textit{Journal of Open Source Software} (JOSS), who provided valuable feedback to improve our \texttt{R} package.
Cattaneo gratefully acknowledges financial support from the
National Science Foundation through grants SES-1947805, DMS-2210561, and SES-2241575, and
from the National Institute of Health (R01 GM072611-16). Jansson gratefully acknowledges financial support from the
National Science Foundation through grant SES-1947662.

\bibliographystyle{plainnat}
\bibliography{CCJM_2025_JOSS-Supplement--bib}
\end{document}